# Confinement–Connectivity Coupling Enables High-Efficiency Piezoionic Transduction

*Tofayel Ahammad Ovee[a], Daniel Kroeger[b], and Jean-François Louf[a,*]*

Tofayel Ahammad Ovee, Jean-François Louf
Chemical Engineering department, Auburn University, Auburn, AL, 36849, USA
Email: jlouf@auburn.edu

Daniel Kroeger
Department of Anatomy, Physiology and Pharmacology, Auburn University, Auburn, AL, 36849, USA

Funding: The cryo-electron microscopy performed in this research was supported by the National Science Foundation (ECCS-2025462) through the Institute for Matter and Systems (IMS). This work was also supported by NSF PoLS 2442203.

Piezoionic hydrogels offer a route to mechanically driven bioelectronic interfaces, but their output is limited by rapid, symmetric ion redistribution that dissipates charge gradients. In biological electrocytes, efficient signal generation arises from the coupling of ion selectivity with spatial confinement that regulates transport. Here, we introduce a confinement–connectivity design strategy for piezoionic hydrogels, implemented through a supramolecular poly(vinyl alcohol)–glycerol–cucurbit[5]uril (PVA–glycerol–CB[5]) mesoporous network with a layered Negative–Neutral–Positive architecture that simultaneously increases pore fraction while reducing characteristic pore size. This architecture constrains ionic redistribution while maintaining a large mobile-ion reservoir, enabling deformation-driven charge separation. Compression generates peak outputs of ~180 mV and ~9 mA and elicits synchronized electromyographic responses in the mouse sciatic nerve without external power. These results establish confinement–connectivity coupling, rather than bulk conductivity, as a materials design framework in which coupling pore connectivity and confinement governs piezoionic transduction.

## 1. Introduction

Mechanically responsive soft materials that convert deformation into electrical signals are central to the development of bioelectronic[1–3] and biointegrated systems[4,5], including neural interfaces[6–8], artificial skins[9–11], and self-powered stimulators[12–14]. Among these materials, piezoionic hydrogels have attracted increasing attention because they transduce mechanical inputs into ionic electrical signals rather than electronic currents, enabling direct compatibility with hydrated and biological environments[15–18]. In contrast to conventional piezoelectric materials, piezoionic systems rely on ion migration driven by mechanical deformation[19], offering intrinsic softness[20], stretchability[21], and biocompatibility[22]. Despite these advantages, most piezoionic hydrogels generate electrical outputs that remain below the thresholds required for direct electrophysiological activation, largely due to uncontrolled bulk ion transport[18].

This limitation reflects a fundamental materials challenge: ion transport in synthetic hydrogels is typically governed by non-selective bulk diffusion[23]. Under mechanical loading, mobile ions rapidly redistribute and dissipate emerging charge gradients, suppressing voltage accumulation and reducing energy conversion efficiency[24]. Biological systems overcome this limitation through spatial organization of ion transport[25]. In electric eel electrocytes, asymmetric membrane permeability is coupled with geometric confinement across thin, compartmentalized domains, which imposes resistance, biases flux directionality, and prevents rapid gradient dissipation[26–28]. This combination of selectivity and confinement enables amplification of weak ionic gradients into coordinated electrical discharges[27,28]. These observations suggest that efficient piezoionic transduction requires not only ion selectivity, but also regulation of transport length scales and redistribution times (**Fig1.a**).

Translating this principle into synthetic materials remains challenging. Most piezoionic hydrogels enhance performance by tuning the properties of the ionic species—such as effective hydrated size[29], valency[32], or concentration[16,30]—to induce asymmetric transport under deformation[31,32], often through differential ionic mobility or drag between cations and anions[29,33] and spatially graded architectures to enhance mechano-ionic coupling[34] . While such strategies can modulate signal magnitude[16], they implicitly treat the polymer network as a passive transport medium and leave the geometry of ion pathways largely unaltered[16].

Consequently, bulk diffusion through percolating pores still dominates[23,24], enabling rapid gradient dissipation and limiting sustained charge separation[35]. An alternative strategy is to engineer the transport geometry itself, such that ionic motion is simultaneously accessible and dynamically constrained. However, how to couple pore connectivity and confinement to regulate ion transport and suppress symmetric redistribution remains largely unexplored.

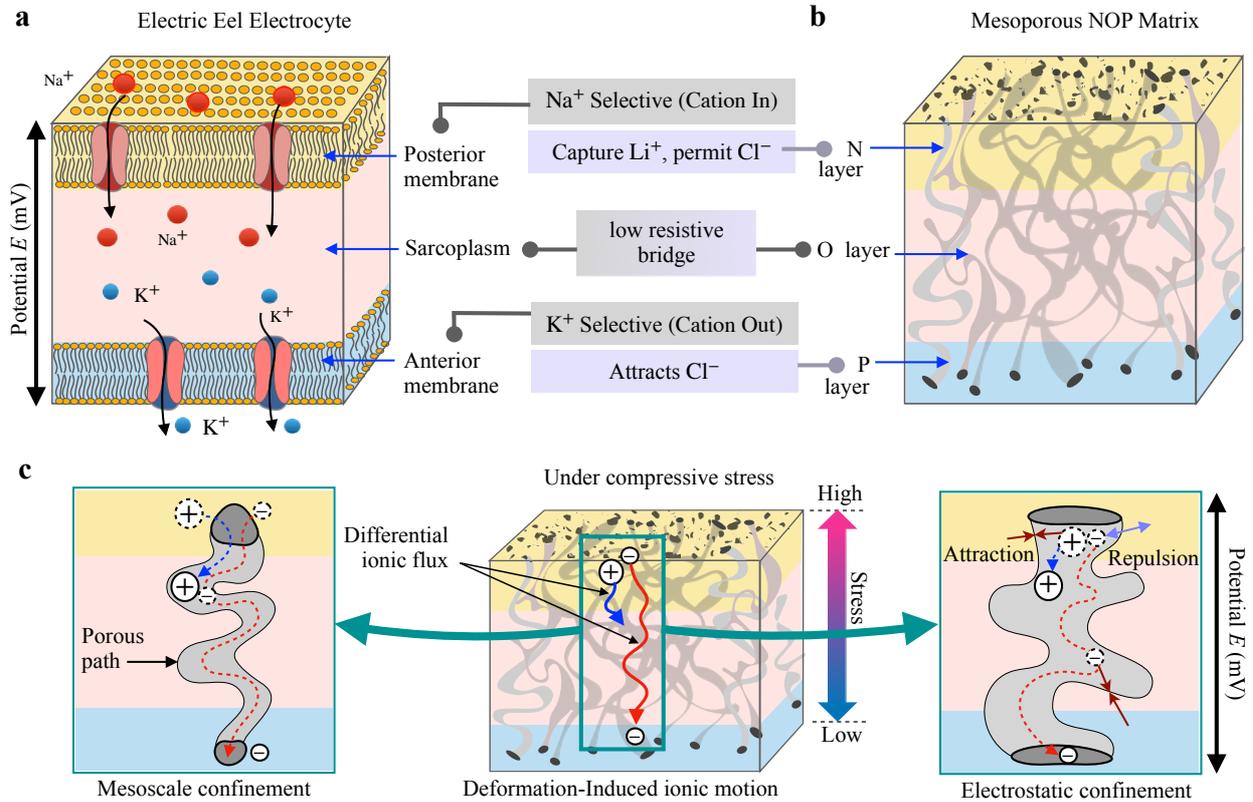

**Figure 1: Conceptual framework for confinement-regulated piezoionic transduction. (a)** In biological electrocytes, asymmetric membrane selectivity combined with geometric confinement biases ion transport and prevents rapid equilibration, enabling voltage generation. **(b)** In the engineered hydrogel, a layered Negative–Neutral–Positive architecture combined with a confined, yet highly connected neutral pore network, enables deformation-driven ion redistribution. Mechanical loading breaks transport symmetry, generating charge separation and electrical output. **(c)** Mesoscale and electrostatic confinement jointly regulate deformation-induced ion transport. Under applied stress, the interconnected porous pathways facilitate differential ionic flux, while local charge domains impose selective attraction and repulsion, leading to asymmetric ion motion and net charge separation across the NOP matrix.

Here, we use confinement–connectivity coupling as a materials design strategy to regulate ion transport in piezoionic hydrogels. By combining a connected mesoporous network with spatially organized charged domains, this architecture is designed to preserve ion accessibility while restricting rapid redistribution. This framework enables us to link pore-scale structure, poroelastic transport, and deformation-induced charge separation to amplified piezoionic output.

We target amplifier-free neural activation as a benchmark for piezoionic transduction, using it to assess whether architecture-controlled ion transport can reach biologically relevant thresholds.

## 2. Results
### 2.1 Construction of the NOP hydrogel architecture

The layered NOP architecture was realized within a supramolecular poly(vinyl alcohol)–glycerol–cucurbit[5]uril (CB[5]) mesoporous network by sequential incorporation of charged polyelectrolytes and host–guest crosslinking (**Fig. 1b**). Poly(sodium 4-styrenesulfonate) (PSSNa) and poly(diallyldimethylammonium chloride) (PDADMAC) were introduced to generate negatively and positively charged domains, respectively, separated by an undoped intermediate layer that maintains mechanical continuity and provides an ionic pathway across the structure. Cryogenic scanning electron microscopy reveals that incorporation of CB[5] significantly alters the internal morphology of the hydrogel (**Fig. 2a–c**), producing a more interconnected mesoporous network compared to PVA and PVA–glycerol controls.

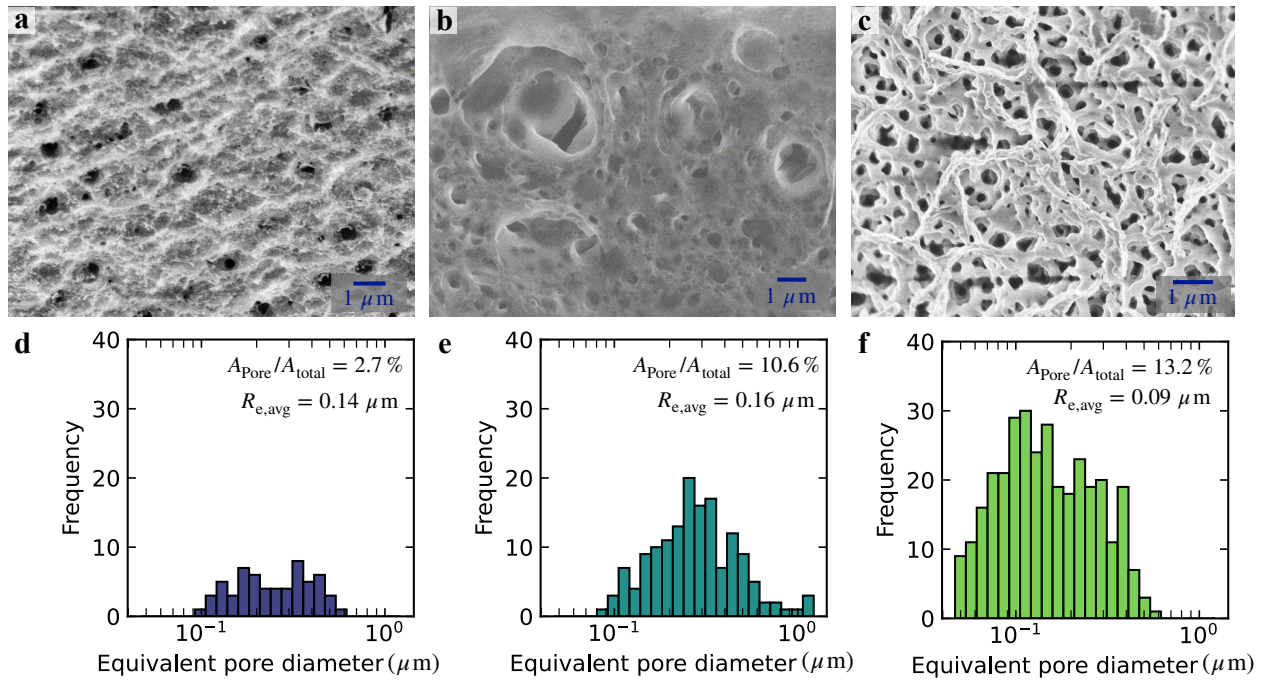

**Figure 2: Pore-scale architecture of PVA-based hydrogels.** (**a–c**) Cryo-SEM images of (**a**) PVA, (**b**) PVA–glycerol, and (**c**) Mesoporous gel (PVA–glycerol–CB[5]), showing distinct

internal morphologies. (**d–f**) Corresponding pore size distributions and equivalent pore radius analysis. Incorporation of CB[5] simultaneously increases pore fraction while reducing the characteristic pore size, producing a network that is both highly connected and spatially confined.

Quantitative image analysis confirms that this structural transition is accompanied by a simultaneous increase in pore fraction and reduction in characteristic pore size (**Fig. 2d–f**). The mesoporous hydrogel network exhibits an areal porosity of 13.2% with a mean pore radius of ≈0.09 μm, compared to 10.6% and ≈0.19 μm for PVA–glycerol, and 2.7% and ≈0.14 μm for PVA. Pore size distributions remain broad, indicating a heterogeneous, multi-scale network that combines high connectivity with spatial confinement.

To decouple these structural changes from bulk mechanical compliance, quasi-static compression tests were performed at low strain rates (**Supplementary Fig. S1**). Despite its higher porosity, the doped mesoporous hydrogel (PVA-glycerol-CB[5]) with NOP configuration exhibits the highest drained modulus ($E_d \approx 82$ kPa) (**Supplementary Section 1**), exceeding both control formulations (**Supplementary Table S1**), indicating that supramolecular crosslinking increases the effective stiffness of the network despite increased porosity, consistent with enhanced load-bearing interactions within a confined pore architecture.

## 2.2 Poroelastic relaxation reveals confinement-regulated transport dynamics

Step compression reveals distinct relaxation dynamics and coupling between mechanical deformation and ionic redistribution (**Fig. 3a**). Force relaxation follows a single-exponential decay[36,37]:

$$F(t) - F_\infty = (F_0 - F_\infty)e^{-\frac{t}{\tau}} \qquad (1)$$

where $F_0$ and $F_\infty$ denote the initial and equilibrium forces, respectively, $t$ is time, and $\tau$ is the characteristic poroelastic relaxation time (**Supplementary Fig. S2**).

PVA exhibits the shortest relaxation time ($\tau = 14.3 \pm 0.1$ s), consistent with relatively unrestricted fluid and ion transport. In contrast, the mesoporous gel (PVA–Glycerol–CB[5]) network displays the longest relaxation time ($\tau = 23.7 \pm 0.1$ s), indicating increased resistance to distribution. Notably, this occurs despite its higher porosity, demonstrating that transport is not limited by pathway availability but by constrained redistribution over reduced length scales.

The doped mesoporous gel with NOP configuration maintains this confinement while introducing charged domains that bias transport asymmetry, resulting in an intermediate relaxation time ($\tau = 19.4 \pm 0.5$ s).

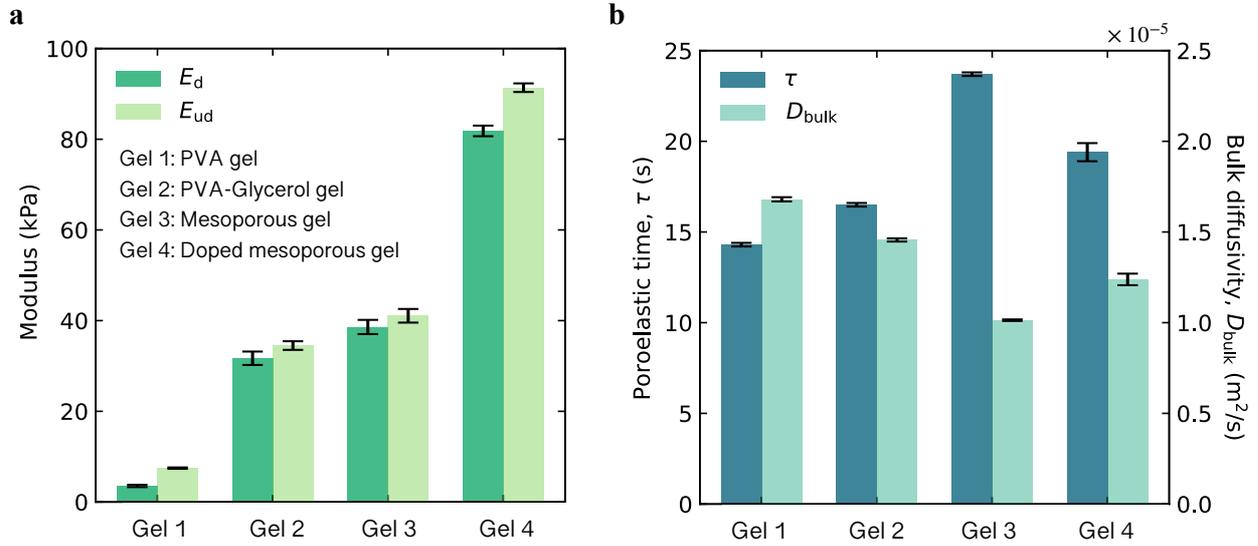

Figure 3. Poroelastic transport dynamics reveal confinement–connectivity coupling. (a) Stress relaxation under step compression for different hydrogel formulations, showing distinct poroelastic timescales. (b) Extracted relaxation time ($\tau$) and estimated bulk diffusivity ($D_{bulk}$). The CB[5]-based network exhibits slower relaxation despite higher porosity, indicating constrained redistribution within a highly connected but spatially confined pore network.

Bulk diffusivity was estimated using $D_{bulk} \approx (L_o - d)^2/\tau$, where $L_o$ is the initial thickness of the gel and $d$ is the applied deformation [38,39]. The doped mesoporous matrix exhibits a moderate diffusivity ($1.24 \times 10^{-5}$ m$^2$s$^{-1}$), confirming that transport timescales are regulated rather than maximized. Rate-dependent modulus measurements further support this interpretation (**Fig. 3b**): the equilibrium modulus ($E_d \approx 82.2$ kPa) is slightly lower than the instantaneous modulus ($E_{ud} \approx 86.8$ kPa) (**Supplementary Fig. S3**), consistent with kinetically constrained fluid redistribution[40]. Together, these results establish that pore-scale organization governs transport through a confinement–connectivity coupling, in which increased pathway availability coexists with reduced redistribution length scales. This regime suppresses rapid equilibration while preserving ionic accessibility, providing the physical basis for enhanced piezoionic transduction in the doped mesoporous hydrogel with NOP configuration.

## 2.3 Confinement–connectivity coupling drives asymmetric ion transport under deformation

The mesoporous (PVA–glycerol–CB[5]) network defines a confined yet highly connected transport environment, while selective doping with PSSNa and PDADMAC generates a trilayer Negative–Neutral–Positive (NOP) configuration with no bulk net charge (**Supplementary Section 2** and **Fig 4a**). Upon equilibration in 4 M LiCl, the high ionic strength screens fixed-charge effects[41,42], and both $Li^+$ and $Cl^-$ infiltrate the network with weak Donnan exclusion[43,44], establishing an approximately electroneutral baseline state (**Supplementary Fig. S4**). This weak electrostatic partitioning was confirmed by measuring the open-circuit membrane potential under symmetric electrolyte conditions, yielding a small potential shift (~7 mV) corresponding to a near-unity ion partition ratio. At this high ionic strength (4 M LiCl), electrostatic interactions are strongly screened, minimizing Donnan exclusion and allowing ions to distribute nearly uniformly throughout the network. These results indicate that, at rest, ionic transport is not governed by static electrostatic gradients (**Supplementary Section 3**).

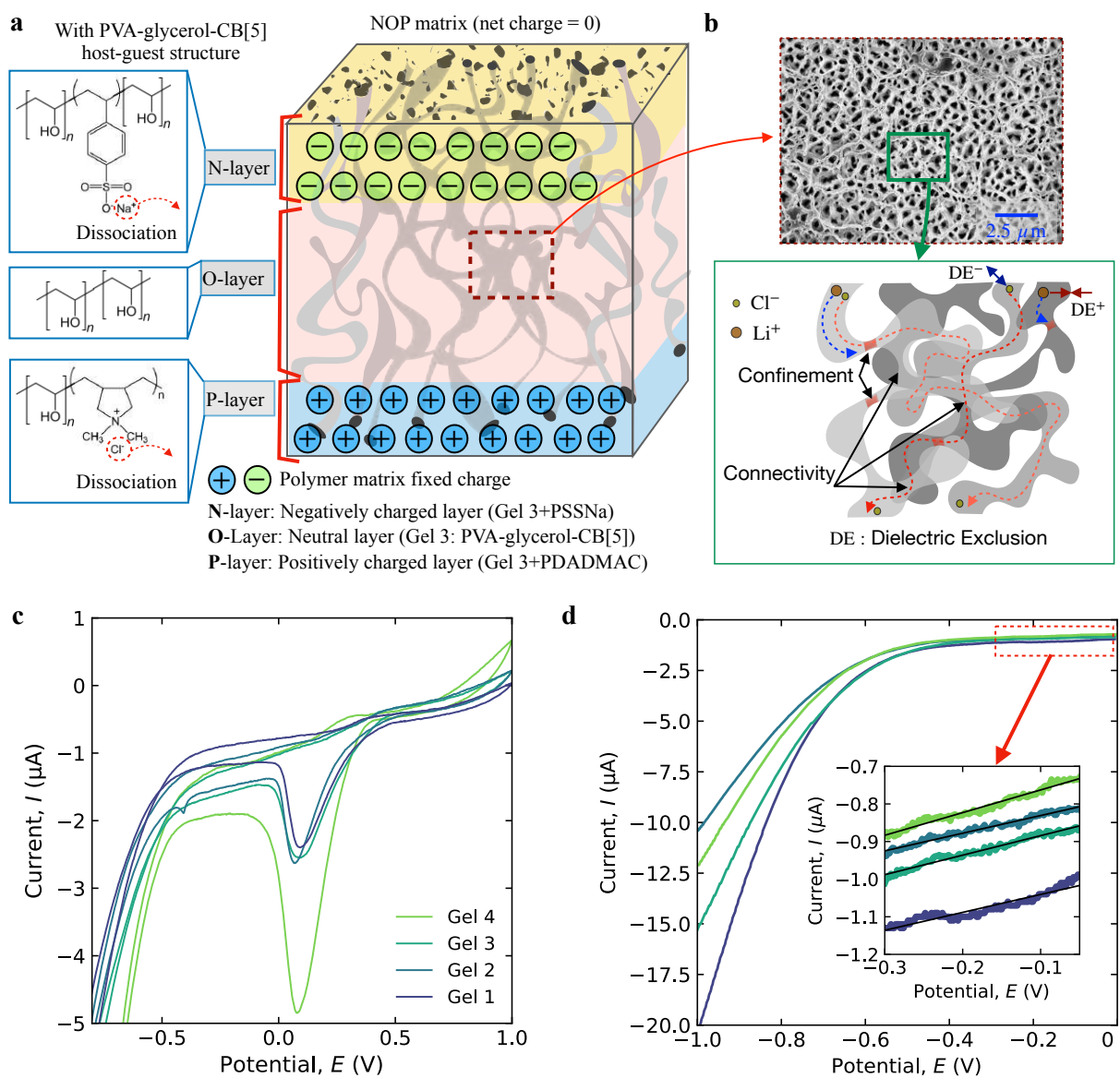

**Figure 4. Deformation-induced asymmetric ion transport in a confined, connected network. (a)** Schematic of the doped mesoporous gel with NOP configuration. **(b)** Mesoscale pore networks impose confinement on hydrated ions within small pores, while in larger pores, dielectric exclusion arising from interactions between charged layers and electrolyte ions governs selective ion mobility, establishing the structural basis for asymmetric transport under deformation. **(c)** Cyclic voltammetry showing enhanced polarization and charge participation in the doped mesoporous matrix. **(d)** Linear sweep stripping voltammetry indicating increased low-bias conductance and field-coupled ionic response.

Under mechanical loading, a stress gradient drives ion migration from compressed to relaxed regions. Within the doped mesoporous NOP configuration, this redistribution becomes asymmetric due to the combined effects of geometric confinement and layered electrostatics: $Cl^-$ (hydrated diameter ≈ 0.66 nm) experiences reduced steric confinement relative to $Li^+$ (≈ 0.76 nm), while electrostatic repulsion from the negatively charged layer and attraction toward the positively charged layer further bias anion transport. The neutral interlayer maintains connectivity while imposing controlled resistance, enabling deformation-induced charge separation across the network (**Fig. 4b**).

Electrochemical measurements provide quantitative signatures of this regulated transport regime. Cyclic voltammetry reveals distinct loop morphologies across hydrogel formulations (**Fig. 4c**): PVA exhibits narrow hysteresis consistent with largely symmetric ionic motion, whereas the doped mesoporous matrix displays broadened, structured profiles (AUC = 1.03 × $10^{-3}$ mA·V) and a pronounced cathodic feature near zero potential, indicative of enhanced polarization under constrained redistribution. Complementary linear sweep stripping voltammetry shows a steeper low-bias conductance slope for the doped mesoporous matrix (dI/dE ≈ 6 × $10^{-4}$ mA·$V^{-1}$; **Fig. 4d**), consistent with increased field-coupled ionic response. Together, these results establish that deformation breaks transport symmetry in a confined yet connected network, converting broadly distributed ionic mobility into directional charge separation and providing the electrochemical basis for amplified piezoionic transduction (**Supplementary Section 4**).

**2.4 Piezoionic output emerges from transport asymmetry under mechanical loading**

For electromechanical measurements, all hydrogel formulations were tested in a pyramidal geometry designed to localize compressive strain and enhance piezoionic output, following prior device-design strategies[45], and performance was evaluated under controlled compression (**Fig. 5a**). The doped mesoporous with NOP configuration generates peak open-circuit voltages of 188 ± 16 mV and short-circuit currents of 8.4 ± 2.8 mA, approximately three orders of magnitude greater than the PVA control (**Fig. 5b**). These outputs fall within the regime of peripheral nerve stimulation, where applied electric fields drive neural activation, motivating direct bio-interfacing tests[46–49]. Under cyclic loading, the doped mesoporous matrix exhibits stable, reproducible voltage and current transients that remain tightly synchronized with deformation and release cycles (**Fig. 5c,d**). The transient response exhibits a characteristic lag (~0.7 s), consistent with poroelastic redistribution dynamics and the measured relaxation times, indicating that electrical output is governed by time-dependent transport within the network (**Supplementary Fig. S5**). The output remains consistent over more than 100 compression cycles with negligible signal attenuation (**Supplementary Fig. S6**), indicating reversible ionic redistribution rather than interfacial degradation or ion depletion.

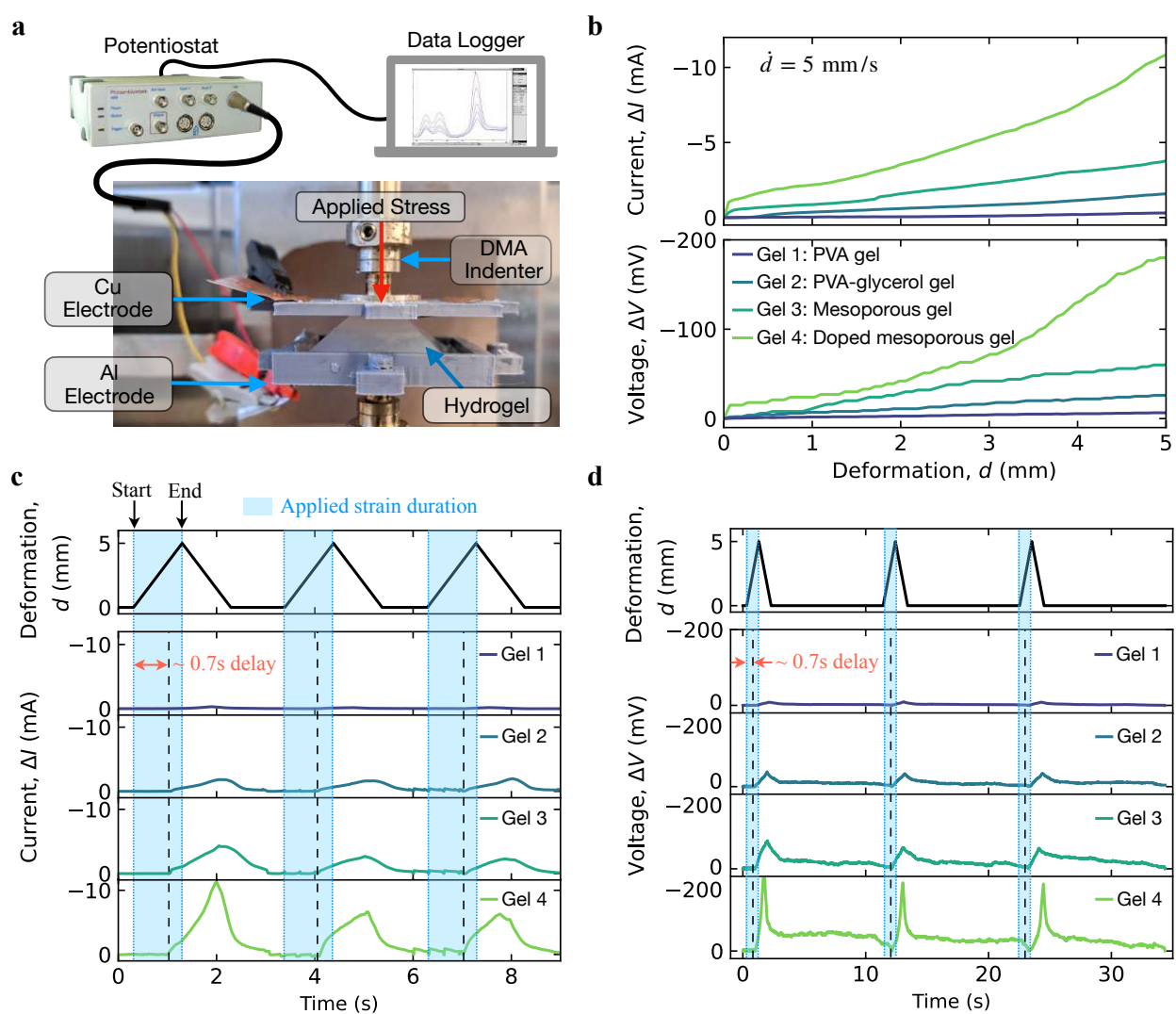

**Figure 5: Piezoionic response under mechanical loading. (a)** Experimental setup combining mechanical compression (DMA) with electrical measurement. **(b)** Peak open-circuit voltage and short-circuit current generated under compression, showing enhanced output for the NOP hydrogel. **(c, d)** Representative voltage and current transients under cyclic loading with a delay time (~0.7s) due to the poroelastic nature of the piezoionic hydrogel, demonstrating a stable and synchronized electromechanical response.

Notably, enhanced output does not scale monotonically with mechanical compliance or pressure sensitivity (**supplementary Section 5**). Although some control hydrogels exhibit greater deformability, they produce lower voltage and power densities. This decoupling indicates that performance is governed by ionic transport under deformation rather than by mechanical response alone (**Supplementary Fig. S7**).

To quantify energy conversion efficiency, we extracted peak voltage, current, and derived power density across hydrogel architectures (**Fig. 6a,b**). The doped mesoporous hydrogel achieves the highest areal power density (7.2 ± 3.1 mW m$^{-2}$), despite only moderate pressure sensitivity. Further insight is provided by transport metrics (**Fig. 6c**): the doped mesoporous matrix simultaneously exhibits the largest transported charge per cycle and the highest low-bias conductance slope, indicating that a greater fraction of mobile ions participates in deformation-coupled redistribution. This divergence between total mobility and usable output demonstrates that mesoscale organization converts broadly distributed ionic motion into directional charge separation. To capture the mesoscale effect, a poroelastic transport factor ($\Pi$) is defined as

$$\Pi \sim \phi \left(\frac{r_a}{r_c}\right)\left(\frac{\tau_p}{t_1}\right) \qquad (2)$$

where $\phi$ is the porosity, $r_c$ is the characteristic pore radius, $r_a$ is the average pore radius, $\tau_p$ is the poroelastic relaxation time, and $t_1$ is the characteristic loading time, defined here as the duration of applied deformation (1 s).

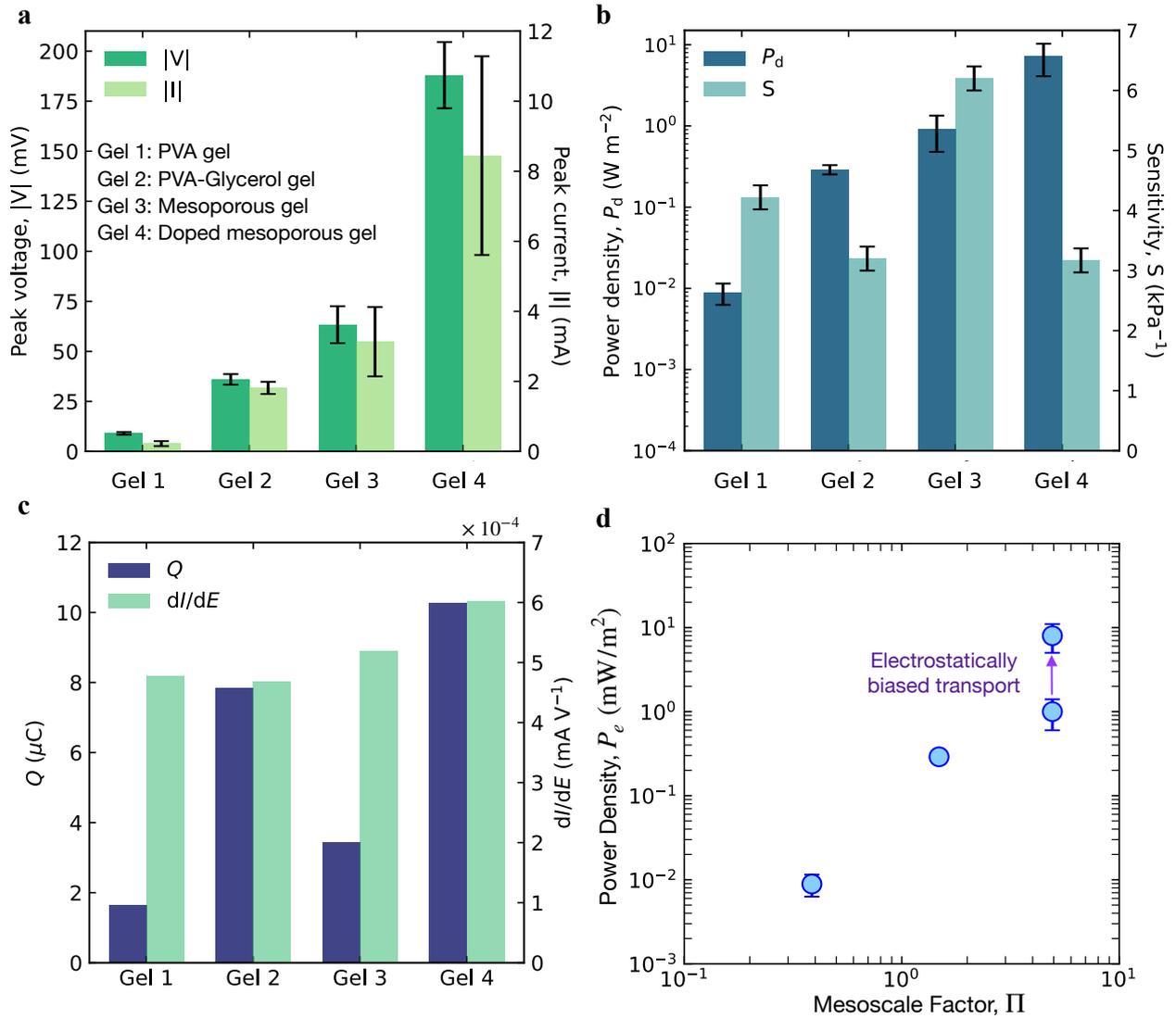

**Figure 6: Transport efficiency and energy conversion across hydrogel architectures. (a)** Peak voltage and current generated under cyclic loading. **(b)** Derived power density and pressure sensitivity, showing that maximum output does not coincide with maximum compliance. **(c)** Transport metrics, including total charge per cycle and low-bias conductance slope, indicate enhanced participation of mobile ions in deformation-driven transport for the NOP architecture. **(d)** Power density $P_e$ as a function of the mesoscale factor $\Pi$, showing a strong positive correlation between structurally governed transport asymmetry and electrical output. For NOP an electrostatically biased transport, amplifies piezoionic power generation.

This factor accounts for the combined effects of pore fraction, confinement length scale, and transport timescale, providing a materials-agnostic parameterization on deformation-induced ionic redistribution. Across the base hydrogel matrices (PVA, PVA–glycerol, and mesoporous CB[5]), $\Pi$ correlates positively with power generation, indicating that increased connectivity combined with constrained transport enhances piezoionic output. Additionally, the NOP configuration enhances the generated power by approximately an order of magnitude, consistent with electrostatically biased transport that amplifies directional ion flux (**Fig. 6d**).

Impedance and load-dependent measurements further support this interpretation. The doped mesoporous hydrogel exhibits elevated bulk resistance ($\approx 4\ \Omega$) and a strongly non-ideal interfacial response ($n \approx 0.24$), consistent with distributed polarization in a confined transport network (**Supplementary Section 6** and **Figs. S8–S10**). Load-dependent measurements reveal a finite internal resistance and a power maximum at intermediate load, confirming that the hydrogel behaves as a regulated ionic source rather than a purely conductive medium (**Supplementary Fig. S11**). Together, these results establish that high piezoionic output arises from transport asymmetry in a confined yet connected network, where confinement amplifies mobility differences associated with hydrated ion size, rather than from increased conductivity or mechanical compliance (**Supplementary Section 7**).

## 2.5 Amplifier-free neural activation enabled by architecture-controlled ion transport

To test whether architecture-controlled transport is sufficient to drive physiological responses, we interfaced the NOP hydrogel with an isolated mouse sciatic nerve (**Fig. 7a**). Upon mechanical actuation, deformation-driven ionic signals generated by the hydrogel were directly delivered to the nerve, eliciting compound muscle action potentials recorded by electromyography (EMG) without external amplification.

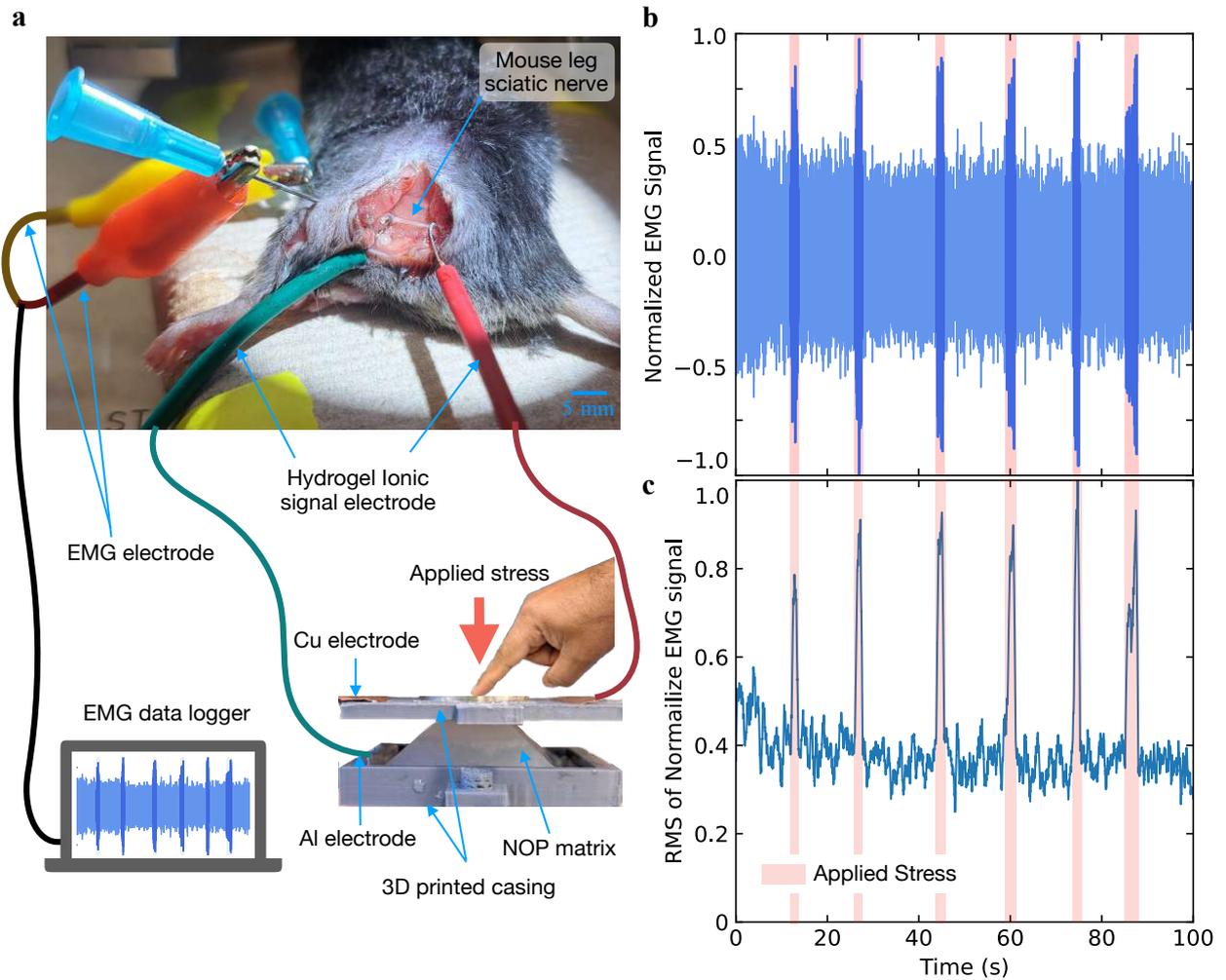

**Figure 7:** *In Vivo* **Validation of Piezoionic Neural Stimulation. (a)** Schematic of the mouse sciatic nerve stimulation model using the doped mesoporous matrix (NOP) as a self-powered electrode. **(b)** Recorded normalized EMG signals triggered by mechanical stress applied to the hydrogel. **(c)** RMS envelope analysis demonstrating stable and reproducible activation exceeding the threshold for motor response.

Repeated compressive loading produced EMG responses tightly synchronized with mechanical input, with consistent amplitudes across cycles (**Fig. 7b**). Root-mean-square (RMS) envelope analysis (**supplementary Section 8**) confirms stable and reproducible activation over prolonged stimulation periods (**Fig. 7c and Supplementary Fig. S12**). These results demonstrate that deformation-induced charge separation within the NOP architecture generates electrical outputs that exceed the activation threshold of peripheral nerve activation leading to motor responses. Importantly, this activation is achieved without increasing bulk conductivity, but through confinement-regulated transport asymmetry that converts mechanical input into usable electrical signals. Together, these results demonstrate that architecture-regulated ion transport can directly transduce mechanical inputs into functional neuromuscular responses, establishing a route to self-powered bioelectronic interfaces.

## 3. Discussion

### 3.1 Confinement–connectivity coupling provides a design rule for piezoionic efficiency

Our results demonstrate that maximizing ionic conductivity alone does not yield efficient piezoionic energy conversion. In conventional hydrogels, rapid and symmetric migration of cations and anions dissipates stress-induced charge gradients, limiting sustained voltage accumulation[50]. By contrast, the doped mesoporous gel with NOP configuration operates in a distinct transport regime characterized by elevated baseline resistance coupled to mechanically activated ion redistribution. Mesoscale confinement within a connected pore network, together with spatially layered fixed charges, suppresses unregulated bulk conduction at rest while promoting directional ionic flux under deformation. This regulated asymmetry enhances charge separation and power generation without increasing mechanical compliance. Together, these results define a materials design rule: high-output piezoionic hydrogels should combine accessible ion reservoirs, confined transport pathways, and spatially biased charge domains rather than maximizing bulk conductivity alone. Although demonstrated here in a PVA–glycerol–CB[5] system, the framework is architecture-driven and is expected to extend to other ionic hydrogel networks in which pore connectivity, confinement length scale, and transport timescales can be independently tuned. This principle parallels biological electrocytes, where membrane resistance and selective gating convert small ionic gradients into functional electrical outputs.

### 3.2 Mesoscale confinement and poroelasticty jointly regulate transport asymmetry

Electrochemical and mechanical measurements indicate that enhanced piezoionic performance arises from regulated ionic redistribution rather than increased total ion flux. In the doped mesoporous matrix, confinement and layered fixed charges bias ion motion under deformation, restricting symmetric conduction at rest while promoting directional transport when stressed. Joint analysis of cumulative charge and low-bias conductance shows that mesoscale organization converts broadly distributed ionic mobility into deformation-coupled charge separation.

This transport asymmetry is further governed by poroelastic dynamics. Hydrogels with rapid relaxation time dissipate stress-induced charge gradients before substantial separation can accumulate, whereas excessively slow transport constrains ionic redistribution[51]. The doped mesoporous matrix operates in an intermediate regime in which transport is sufficiently restricted to sustain voltage buildup while remaining responsive to mechanical input. This balance defines a temporal window for electromechanical transduction, linking transport timescale to output efficiency.

### 3.3 Baseline electrochemical conditioning enables mechanically gated transduction

The elevated baseline resistance and distributed phase response of the doped mesoporous hydrogel indicate a constrained ionic state in which mobile ions are present but not freely conducting. Mechanical deformation perturbs this state by modifying pore geometry and electrostatic gradients, thereby activating directional ion redistribution across the layered structure. This transition from a constrained to an active transport state enables large transient voltage and current responses. Unlike continuously conductive hydrogels, in which deformation primarily redistributes charge symmetrically[52], the NOP architecture couples mechanical gating to transport asymmetry, enabling large piezoionic outputs without external amplification.

### 3.4 Implications for bioelectronic interfaces

The ability of the doped mesoporous hydrogel to induce reproducible peripheral nerve activation and motor responses demonstrates the functional relevance of confinement-regulated ion transport for soft bioelectronic systems. Neural activation arises not from increased bulk conductivity, but from deformation-induced charge separation governed by spatial and temporal control of ionic motion. More broadly, these results establish general design principles for high-performance piezoionic materials: (i) ion selectivity must be coupled to structural confinement to sustain voltage buildup; (ii) hierarchical charge organization enables directional transport without static electrostatic bias; and (iii) poroelastic transport sets the temporal window for efficient electromechanical coupling. Electrochemical metrics such as cumulative charge and low-bias conductance provide predictive indicators of transduction performance. Together, these findings define a framework in which architecture, rather than composition alone, governs ionic transport and enables self-powered bioelectronic function.

## 4. Methods
### 4.1 Materials

Poly(vinyl alcohol) (PVA, 98–99% hydrolyzed, medium MW; Thermo Scientific), cucurbit[5]uril (CB[5]) hydrate (Sigma-Aldrich), poly(sodium 4-styrenesulfonate) (PSSNa, average MW 70 kDa; Thermo Scientific), poly(diallyldimethylammonium chloride) (PDADMAC, high MW; Sigma-Aldrich), glycerol (VWR), and lithium chloride (LiCl, anhydrous, ACS reagent; Sigma-Aldrich) were used as received. Deionized water (18.2 MΩ·cm) was used for all solutions. PLA filament (eufymake) was used for 3D-printed molds.

### 4.2 Fabrication of layered supramolecular hydrogels

All hydrogels were prepared under heating (130 °C) with continuous stirring (300-400 rpm) and cast into 3D-printed PLA molds, followed by freeze–thaw processing. Unless stated otherwise, samples were equilibrated in 4 M LiCl for 12 h before testing (**Supplementary Table S2**). PVA hydrogels were prepared by dissolving 3 g PVA in 36 g water (~7.7 wt%). After casting, samples underwent one freeze–thaw cycle (−20 °C, 12 h; 25 °C, 2 h) to induce physical crosslinking. PVA–glycerol hydrogels were prepared similarly using a mixed solvent of 20 g water and 16 g glycerol. Supramolecular PVA-Glycerol-CB[5] hydrogels were prepared by dissolving 3 g PVA in 18 g water and 16 g glycerol, followed by addition of 2 mL CB[5] solution (1 mg mL$^{-1}$). The Negative–Neutral–Positive (NOP) matrix was assembled as a layered supramolecular PVA–Glycerol-CB[5] construct. Each layer used a base formulation of 3 g PVA, 16 g glycerol, and 2 mL CB[5] solution (1 mg mL$^{-1}$). The negatively charged (N) layer was formed by adding 1 mL of 1 M PSSNa to 17 g water; the positively charged (P) layer by adding 1 mL of 1 M PDADMAC under identical conditions. The neutral (O) layer contained no added polyelectrolyte and used 18 g water. Layers were cast and frozen at −20 °C sequentially in the mold: N layer (6 hrs), O layer (6 hrs), and P layer (12 hrs), enabling interfacial bonding while preserving layer-specific electrochemical properties. The assembled NOP hydrogel was equilibrated (12 hrs) in 4 M LiCl prior to characterization.

The NOP matrix has a top surface area $A_t = 15 \times 15 \text{mm}^2$ and a bottom surface area $A_b = 25 \times 25 \text{ mm}^2$ with a total thickness of 17 mm. Both the N and P layers were fabricated using 1 mL of 1 M PSSNa and PDADMAC, respectively, within a PVA–CB[5] matrix (**Supplementary Table S2**). A top N-layer thickness of $l_N = 5 \text{mm}$ and a bottom P-layer thickness of $l_P \approx 2.3 \text{mm}$ ensure electroneutrality, yielding equal and opposite fixed charges ($Q_N \approx -96.5$ C, $Q_P \approx +96.5$ C; **Supplementary Section 2**).

### 4.2 Structural characterization

Internal morphology was examined by cryogenic scanning electron microscopy (cryo-SEM). Cryo-SEM was performed using a Thermo Fisher Helios 5CX focused ion beam scanning electron microscope (FIB-SEM). The sample was initially frozen in liquid nitrogen to preserve its native water content, then transferred to a transfer chamber attached to the SEM chamber. Surface water was removed by sublimation at −100 °C for 5 mins, followed by sputter coating with a platinum layer approximately 20–30 nm thick. The sample was transferred to the SEM chamber and maintained at −140 °C during imaging. Pore size distributions and area-based porosity were quantified from Cryo-SEM images using threshold-based image. Due to the complex porous structure, grayscale thresholding was performed using FIJI (ImageJ), and pore boundaries were further validated through manual outlining. The resulting segmented images were then processed using Python to extract pore size distributions and quantitative morphological parameters.

### 4.3 Mechanical and poroelastic characterization

Poroelastic relaxation was measured using a dynamic mechanical analyzer (DMA RSA3, TA Instruments) with a flat-ended indenter. Samples were subjected to stepwise compressive deformation at a constant displacement rate followed by hold periods. Normalized force relaxation curves were fitted to a single-exponential decay to extract poroelastic relaxation times ($\tau$). The effective compressive modulus was obtained from stress–strain measurements performed at slow and fast deformation rates to determine equilibrium and instantaneous responses, respectively (**Supplementary Section 1**).

### 4.4 Electrochemical characterization

*4.4.1 Baseline measurements.* Electrical resistance ($R$), capacitance ($C$), impedance ($Z$), and phase angle ($\theta$) were measured at zero deformation using a Matrix MCR-5100. A moderate fixed frequency of 1 kHz was used for all baseline measurements, chosen to minimize low-frequency interfacial polarization effects while maintaining stable signal acquisition[51]. Measurements were performed in a two-electrode configuration with aluminum (contact area $25 \times 25$ mm$^2$) and copper electrodes $15 \times 15$ mm$^2$ in direct contact with hydrogel (only gravitational loading).

*4.4.2 Dynamic response under deformation.* Electrochemical parameters (R, C, Z, θ) were recorded during compression–release cycles by integrating the hydrogel into the DMA setup coupled to the Matrix MCR-5100 (1 kHz).

*4.4.3 Cyclic voltammetry and linear sweep analysis.* Electrochemical measurements were performed using an eDAQ EA163 potentiostat under identical conditions for all samples in a three-electrode configuration[53], where the working electrode was connected to the Cu electrode, and both the reference and ground electrodes were connected to the Al electrode. Staircase cyclic voltammetry was conducted over a ±1 V window at a scan rate of 100 mV s$^{-1}$ (step height 2 mV, step width 20 ms; 2000 steps) following a 2 s rest period[54]. Linear sweep stripping voltammetry was performed from −1000 mV to 0 mV at 100 mV s$^{-1}$ (500 steps) after a 60 s deposition at −1000 mV and a 2 s rest period[55]. Conductance was extracted from linear fits in the low-bias regime.

## 4.5 Piezoionic signal measurements

Piezoionic open circuit voltage ($V$) and short-circuit current ($I$) were recorded (100 points/s) during controlled compressive deformation using the DMA RSA3 synchronized with the eDAQ EA163 potentiostat and *Chart* acquisition software. Peak values were extracted from time-resolved traces aligned with mechanical loading.

## 4.6 EMG acquisition and signal processing

*4.6.1 Animal preparation.* Experiments were performed using 8–12-week-old C57BL/6J mice (Jackson Laboratories, strain #000664; number of mice = 3). Mice were deeply anesthetized with 4% isoflurane and euthanized via cervical dislocation. Measurements were conducted immediately post-mortem to assess femoral nerve activation and lasted <15 minutes. Based on preliminary tests using a standard power supply, a functional time window of approximately 25 minutes was identified, during which reproducible EMG responses could be recorded before signal degradation. The femoral nerve on the left hind limb was surgically exposed and kept moist with sterile saline.

*4.6.2 Experimental setup.* The hydrogel device was interfaced with Al and Cu electrodes, which were connected through jumper cables to the exposed nerve (approximately 5mm apart), serving as the piezoionic signal input pathway. Muscle activation was recorded by inserting two 22G needle electrodes approximately 1 cm apart into the quadriceps femoris. These electrodes were connected via shielded cables to an acquisition system, and EMG signals were digitized and recorded using SleepSign for Animals software (Kissei Komtec Co., Ltd., Japan). The setup assures no direct electrical contact between stimulation and recording circuits. This configuration minimized electrical artifacts and allowed accurate measurement of neuromuscular responses induced by the hydrogel.

*4.6.3 Data processing.* Electromyography (EMG) signals were band-pass filtered using a 4th-order Butterworth filter (50–70 Hz) for mouse model 1 and or (20-100 Hz) for mouse model 2 and 3 with zero-phase forward–backward filtering[56,57]. Muscle activation was quantified using a root-mean-square (RMS) envelope computed with a 0.5 s sliding window[58]. Triggered events were identified from local maxima with a minimum inter-peak separation of 2 s and quantified relative to a 1 s pre-trigger baseline; amplitudes were normalized for visualization.

## 4.7 Data analysis and Reproducibility

All experiments were performed on independently prepared hydrogel samples. Measurements were repeated at least 3 times under identical conditions, and the results were averaged to obtain a single representative value per sample. Data are reported as mean ± standard deviation unless otherwise stated. Curve fitting and data analysis were performed using Python. Nonlinear least-squares fitting was used for model-based analysis (*e.g.,* poroelastic relaxation and electrical response). The doped mesoporous (PVA-GLYCEROL-CB[5]) matrix with (NOP) configuration was reproducibly fabricated over 10 batches, yielding consistent mechanical, electrochemical, and piezoionic performance.

## Acknowledgements

The authors thank the Center for Polymer and Advanced Composites (CPAC), Auburn University, and Dr. Ramsis Farag for providing research facilities. The authors also thank the Institute for Matter and Systems (IMS), a member of the National Nanotechnology Coordinated Infrastructure (NNCI) at Georgia Tech, for Cryo-SEM sample preparation and imaging.

## Declarations

**Competing interests**

The authors declare no conflict of interest.

**Ethics approval and consent to participate**

Not applicable.

**Consent for publication**

Not applicable.

**Data availability**

All data supporting the findings of this study are provided with the submission.

**Materials availability**

Not applicable.

**Code availability**

Code link shared through Zenodo (see link below).

**Author contribution**

Tofayel Ahammad Ovee: Conceptualization, Methodology, Data curation, Writing – original draft, Visualization.

Daniel Kroeger: Bio sample preparation, Experiment, Writing – review & editing.

Jean-François Louf: Supervision, Methodology, Conceptualization, Funding acquisition, Visualization, Writing – review & editing.

**AI use disclosure**

The authors used AI-assisted tools [ChatGPT (OpenAI), Gemini (Google)] for language editing and refinement of the manuscript. All scientific content, analysis, and conclusions were developed and verified by the authors.

**Data Availability Statement**

Data and source code is available through private link:

https://zenodo.org/records/19893453?preview=1&token=eyJhbGciOiJIUzUxMiJ9.eyJpZCI6ImMzNTRkY2Y0LTU1YTgtNDgwZS05ZDc2LThjOTJiZDRmNTc0ZCIsImRhdGEiOnt9LCJyYW5kb20iOiIzMDFhNWUwZmI3NmJhZWU0YTM5ODJkZGUwMDJkMTZkNiJ9.mxP7jwudCZCZ0RJ1t9RkXXP0JpLeBxSRyg-7RFkItjFVNl6K3cBEMBoqmL5fcf1am2K5Hus_6UOHbCB9291uLg